\documentclass[useAMS,amssymb,epsfig,usegraphicx,usenatbib,twocolumn,revtex4-1]{emulateapj}

\def\araa{ARA\&A}
\def\apj{ApJ}
\def\apjl{ApJ}
\def\apjs{ApJS}
\def\aap{A\&A}
\def\mnras{MNRAS}
\def\nat{Nature}
\def\na{NewA}

\def\aaps{A\&AS} 

\usepackage[usenames,dvips]{color}

\newif\ifAMStwofonts

\makeatother

\shorttitle{Lost in secular evolution}
\shortauthors{Saha}
\begin{document}
\title{Lost in secular evolution: the case of a low mass classical bulge}

\author {Kanak Saha}
\affil{Inter-University Centre for Astronomy and Astrophysics, Pune-411007, India, \\e-mail: kanak@iucaa.ernet.in}

\label{firstpage}

\begin{abstract}

The existence of a classical bulge in disk galaxies holds important clue to the assembly 
history of galaxies. Finding observational evidence of very low mass classical bulges 
particularly in barred galaxies including our Milky Way, is a 
challenging task as the bar driven secular evolution might bring significant dynamical 
change to these bulges alongside the stellar disk. 

Using high-resolution N-body simulation, we show that if a cool stellar 
disk is assembled around a non-rotating low-mass classical bulge, the disk rapidly grows a 
strong bar within a few rotation time scales. Later, the bar driven secular process transform 
the initial classical bulge into a flattened rotating stellar system whose central part
also have grown a bar-like component rotating in sync with the disk bar. During this time, 
a boxy/peanut (hereafter, B/P) bulge is formed via the buckling instability of the disk bar and 
the vertical extent of this B/P bulge being slightly higher than that of the classical 
bulge, it encompasses the whole classical bulge. 
The resulting composite bulge appears to be both photometrically and kinematically identical 
to a B/P bulge without any obvious signature of the classical component. Our analysis suggest that
many barred galaxies in the local universe might be hiding such low-mass classical bulges.
We suggest that stellar population and chemodynamical analysis might be required in 
establishing the evidence for such low-mass classical bulges.
\end{abstract}


\section{Introduction}
\label{sec:intro}
Classical bulges are typically thought to have formed as a result of violent 
mergers \citep{Kauffmanetal1993, Baughetal1996, Hopkinsetal2009}  
or collapse of primordial gas clouds \citep{Eggenetal1962}  or coalescence of giant clumps 
in high-redshift galaxies \citep{Immelietal2004, Elmegreenetal2008} or multiple minor 
mergers \citep{Bournaudetal2007, Hopkinsetal2010}.
Size and mass of these classical bulges depend
on the process that formed these stellar system but perhaps little on the subsequent 
evolution. Observation suggests that the classical bulges are generally dispersion 
dominated, spheroidal systems which might remain dynamically unchanged over 
several billion years. A number of recent studies suggest coexistence of classical bulges
and bars in disk galaxies \citep{Gadotti2009,Erwinetal2015}. Of particular interest are
the low-mass classical bulges in barred galaxies, such as our own Galaxy.
N-body modelling of BRAVA stellar kinematics reveals that our Milky Way might 
harbor a low-mass classical bulge inside it's B/P bulge \cite{Shenetal2010,DiMatteoetal2015}. 
For a current summary of the Milky Way's bulge, the readers are referred to \cite{Gerhard2014}.
Recent simulation by \cite{Sahaetal2012} showed that such a low-mass classical bulge is 
significantly modified during the secular evolution along with the cool stellar disk. 
It remains unclear whether such low-mass classical bulges can reliably be detected in observation. 

In the initial phase of galaxy evolution, disk galaxies (at high redshift) go through 
violent instabilities \citep{Genzeletal2006} and possibly have formed various non-axisymmetric 
structures (most prominent of which are bar and spiral arms) which, later, drive 
the slow secular evolution of these galaxies. In the secular phase, the most
efficient way a disk galaxy evolves is through forming a bar which facilitates the 
redistribution of energy and angular momentum between the disk, dark matter halo and 
the preexisting classical bulge \citep{DebattistaSellwood2000,Athanassoula2003, Sahaetal2012}.
As the bar becomes stronger, it goes through buckling instability and form
boxy/peanut bulges as demonstrated in numerous N-body simulation
studies \citep{CombesSanders1981,PfennigerNorman1990, Rahaetal1991, MV2004,
Sahaetal2012}. More than $40$\% of all observed bulges are B/P bulges 
\citep{Luttickeetal2000a} and their observed properties are nicely summarized in 
\cite{LaurikainenSalo2015}. The formation of a B/P bulge 
in the central region is an energetically favorable phenomenon, as the B/P bulge is 
comparatively more random motion dominated than the initial disk. In essence, secular 
evolution transforms the central part of a 
cool stellar disk to one with a comparatively hot centrally concentrated bulge \citep{KK2004}.
When such a process takes place in the presence of an 
initially non-rotating low-mass classical bulge at the center, the secular process 
also transforms the low-mass classical bulge to a fast rotating object whose kinematic 
properties resemble that of a B/P bulge, namely having cylindrical rotation in the inner
region \citep{Sahaetal2012}. The final composite bulge is a superposition of the 
B/P bulge and the transformed classical bulge, both of which are cylindrically rotating.
The goal of this paper is to find out whether such a low-mass classical bulge can
be unmasked from the composite bulge. We present a systematic morphological and kinematic 
analysis to extract information about the presence of the classical bulge in the barred galaxy. 

\begin{figure*}
\begin{flushleft}
\rotatebox{0}{\includegraphics[height=8.5 cm]{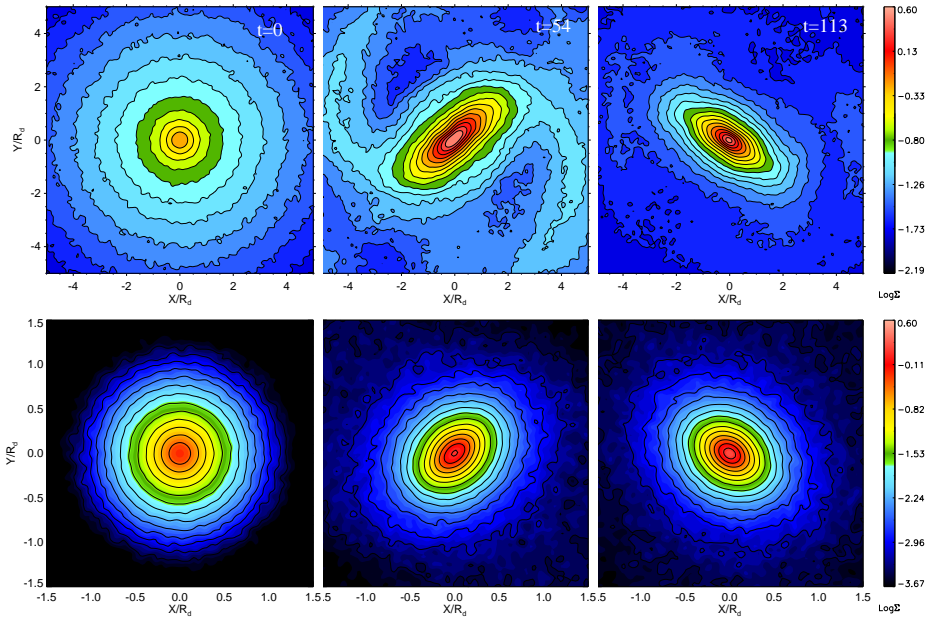}}
\caption{Upper panel: face-on stellar surface density (including both disk stars as well 
as classical bulge stars) maps in model RCG004 at t=0, 54 and 113 
(internal unit). Bottom panel: face-on surface density maps of the classical bulge stars alone at 
same epochs.}
\label{fig:contdenmap}
\end{flushleft}
\end{figure*}
  
\section{galaxy model and simulation}
\label{sec:modelsetup}
The initial galaxy model, in equilibrium, consists of an initially axisymmetric 
exponential disk, a cored dark matter halo and a non-rotating classical bulge. 
The initial classical bulge (hereafter, ICB) is strongly flattened by the disk gravity. 
The ICB in this galaxy model has a total mass of $M_b = 0.06 M_d$, $M_d$ being the
disk mass, and initial ellipticity in edge-on projection is given by $\epsilon_b = 0.46$.
The initial stellar disk is cool with Toomre $Q =1.4$ at $R=2.5 R_d$.
Further details on the model is given in \cite{Sahaetal2012}.
  
We scale the model such that $M_d = 4.58 \times 10^{10} M_{\odot}$ and $R_d
=4.0$ kpc. Then the time unit is given by $24.9$ Myr. We have used
 a total of $10$ million particles. The softening lengths for the disk, 
bulge and halo particles used are $12$, $5$ and $20$ pc respectively 
following the suggestion of \cite{McMillan2007}. The simulation is performed
using the Gadget code \citep{Springeletal2001} with a tolerance parameter
$\theta_{tol} =0.7$, integration time step $\sim 0.4$ Myr. The simulation 
was evolved for a time period of $\sim 3.0$ Gyr.

\section{Morphology of the composite bulge}
\label{sec:boxy}
The initial axisymmetric stellar disk, being cool, rapidly
forms a bar with two-armed spiral as shown in Fig.~\ref{fig:contdenmap}. The spiral
produces radial heating \citep{Sahaetal2010,SellwoodCarlberg2014}, which eventually 
destroys itself leaving a strongly barred galaxy at the end of the simulation. 
The end product resembles a typical barred lenticular galaxy without any spiral 
arm \citep{Cortesietal2013}. In the center of this barred galaxy lies the low-mass 
classical bulge (see rounder contours at $t=0$ in Fig.~\ref{fig:contdenmap}). 
However, as time progresses, even at the very central 
region ($< 0.5 R_d$) the density contours are no longer rounder but are purely elliptical 
suggesting non-existence of the classical bulge component. We have checked this further 
inside $R < 0.2 R_d$ and found similar result. In fact, the face-on surface 
density maps of the classical bulge stars 
alone (bottom panel of Fig.~\ref{fig:contdenmap}) show that the inner part of this 
bulge is actually a bar - together with the disk stars, the central part of the galaxy 
behaves like a single bar component without any structural signature of the classical bulge. 
It has been shown previously that such a classical bulge actually absorbs a significant 
fraction of the disk angular momenta emitted by the bar and is transformed into a 
rapidly rotating bar-like object \citep{Sahaetal2012}. 

\begin{figure}[b]
\rotatebox{270}{\includegraphics[height=8.0 cm]{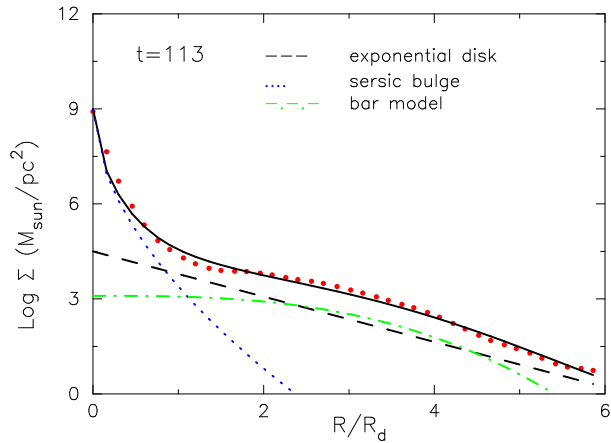}}
\caption{Radial surface density profile of the barred galaxy at $t=113$ and its 
decomposition into a bulge+bar+disk. The sersic index of the bulge is
$n=1.82$ and the bar is modelled with a sersic index $n=0.35$.  }
\label{fig:radialprofile}
\end{figure}

\begin{figure*}
\rotatebox{0}{\includegraphics[height=7.0 cm]{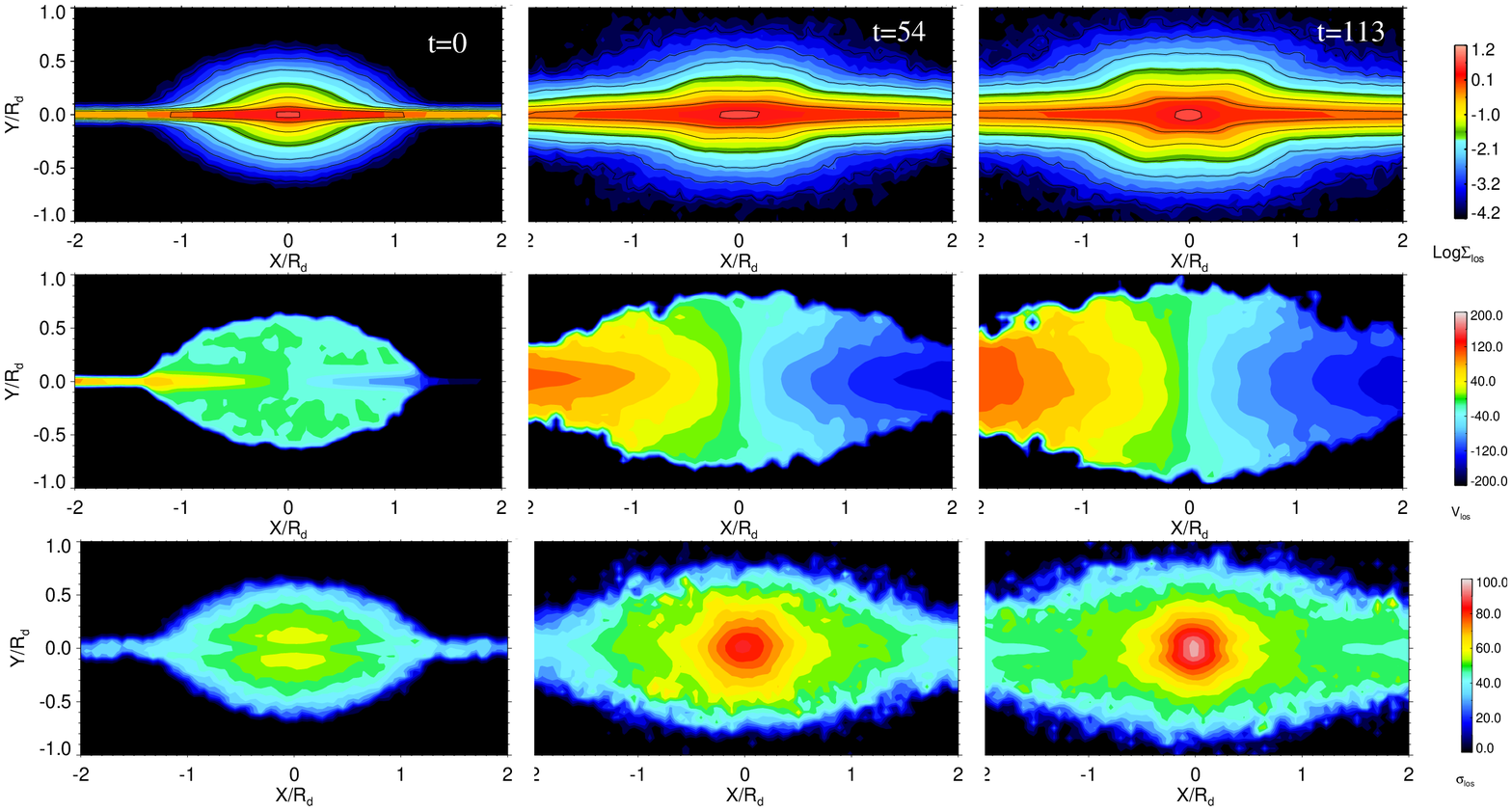}}
\caption{2D line-of-sight moment maps (edge-on projection with the major axis of the 
bar aligned parallel to x-axis) at three different epochs during the evolution.
The upper panel denotes surface density, middle panel the mean velocity and the 
lower panel the velocity dispersion.}
\label{fig:edgeonmaps}
\end{figure*}

In edge-on projection, as shown in Fig.~\ref{fig:edgeonmaps}, the galaxy model at $t=0$
has a classical bulge which extends upto about $0.5 R_d$ in the vertical direction along
the minor axis. As the bar undergoes buckling instability, it forms a B/P bulge
in the inner part of the disk. The outcome of this physical process is a composite bulge
which is a superposition of the preexisting classical bulge and the B/P bulge. Such composite
bulges are not uncommon and they are reported in a number of barred galaxies, recently 
studied by \cite{Gadotti2009, Erwinetal2015}.
The vertical extent of the B/P bulge is about $0.7 R_d$ (see Fig.~\ref{fig:edgeonmaps})
which completely encompass the flattened classical bulge. In other words, the classical
bulge camouflages inside the B/P bulge. This may surprise some readers; a priori it is
not clear what would be the vertical extent of a B/P bulge with a given size and mass of
a classical bulge to begin with the initial disk. In any case, the generality remains unclear 
at the moment, as more such N-body experiments are needed. Obviously, massive and bigger
classical bulges can be excluded as those can not remain immersed in the B/P bulge. 
Note the surface density maps of these composite bulges, shown in Fig.~\ref{fig:edgeonmaps},
appear identical to a pure B/P bulge with no morphological signature of the low-mass 
classical bulge. In other words, the present study suggests that a B/P bulge in a galaxy
might hide a low-mass classical bulge which carries valuable information on the galaxy 
formation history. 

To confirm, we analyzed the surface density profiles at different epochs during the
secular evolution of this galaxy model. Fig.~\ref{fig:radialprofile} depicts the density profile
taken at $t=113$. We have performed 3-component decomposition on this barred galaxy model: 
an exponential stellar disk, sersic bulge and a sersic bar model \cite{Gadotti2009}.
Our three component decomposition shows that the central region can be well modelled
by a pure sersic bulge with sersic index $n=1.8$ without a need for a classical bulge, 
atleast in photometric sense. The radius at which the bulge and disk densities are equal 
is given by $R_{bd}=0.83 R_d$. It appears that the final barred galaxy contains no 
observable morphological evidence of a classical bulge in the bulge dominated region ($R < R_{bd}$). 
In the next section, we analyze the kinematics of this composite bulge. 

\section{Kinematics of the composite bulge}
\label{sec:bkin}
Classical bulges are kinematically hotter with rotational motion
intermediate between the ellipticals and B/P bulges \citep{KI1982}.
As per kinematics concerned, B/P bulges are further distinguished 
from the classical bulges, in that B/P bulges possess cylindrical 
rotation \citep{K1982}, but see \cite{Williamsetal2011, SahaGerhard2013} 
for exceptional cases. The kinematics of B/P structures including 
various projection effects in simulation of disk galaxies are discussed
in depth by \cite{Debattistaetal2005,IannuzziAthana2015}. 

\begin{figure}
\rotatebox{270}{\includegraphics[height=8.8 cm]{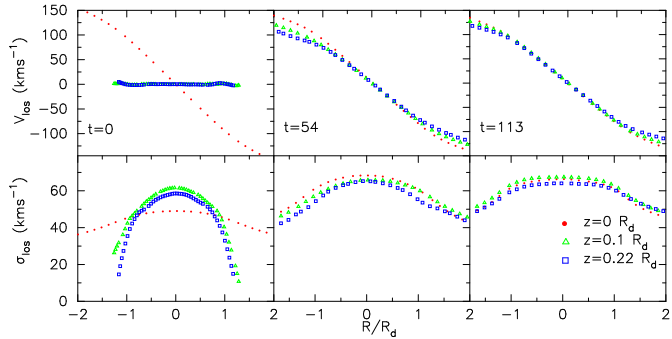}}
\caption{Long-slit stellar kinematics from the region covering the 
composite bulge: the upper panel shows 
the line-of-sight stellar velocity profiles at 3 different heights above the disk 
mid-plane during the evolution. The lower panel shows the same but for the velocity 
dispersion profiles. Note that at $t=0$, the slit at $z/R_d = 0.1$ and $0.22$ 
essentially probe the classical bulge stars.}
\label{fig:slitkin}
\end{figure}

Fig.~\ref{fig:edgeonmaps} depicts the 2D line-of-sight kinematic map 
(in edge-on projection) e.g., mean velocity and velocity dispersion, 
before and after the formation of the B/P bulge in our model galaxy.
While creating these maps, we have rotated the bar so that its major axis
is aligned with the $x$-axis and restricted to one projection only.
Clearly, the ICB is non-rotating. Once the B/P bulge is 
formed and evolved, the final composite bulge (classical + B/P bulge) shows clear 
cylindrical rotation. In addition to the 2D kinematic map, we check this using 
long-slit kinematics at three different heights above the disk mid-plane ($z=0$) 
taken at three different epochs during the secular evolution, see Fig.~\ref{fig:slitkin}. 
It is evident from the figure that the classical bulge has a $V/\sigma =0$ at $t=0$.
The disk rotation velocity rises outward and the velocity dispersion remains nearly flat
in the inner region, as a result the local $V/\sigma$ rises linearly outward.
Note that the local $V/\sigma$ for the slit-1 ($z/R_d = 0$) goes to zero at $R=0$ and this
basically probes the differentially rotating stellar disk. At subsequent times, we have calculated 
the local $V/\sigma$ at $R_{bd} = 0.83 R_d$
and in all cases, we found $V/\sigma > 1.0$ suggesting a kinematically cool bulge.
At $t = 54$, the composite bulge rotates cylindrically. The degree of cylindrical rotation becomes stronger with times \citep[as shown by][]{SahaGerhard2013}, see the panel at $t = 113$ of 
Fig.~\ref{fig:slitkin}.  
The major axis velocity dispersion is nearly flat within the extent of the B/P bulge or even
within $R_{bd}=0.83 R_d$ and exhibit a shallow decline beyond that. The local $V/\sigma$ at 
$R_{bd}$ for the composite bulge at $t=113$ is close to $1.1$ which can be considered as 
a kinematically cool component \citep{Erwinetal2015}. But even otherwise, the kinematic 
signatures are definitely not of a classical component.
So far all the kinematic diagnostics suggest that this composite bulge is actually a 
B/P bulge both morphologically and kinematically. It remains puzzling how to separate 
classical bulge stars that are hidden inside the B/P bulge.          

\begin{figure}
\rotatebox{0}{\includegraphics[height=6.7 cm]{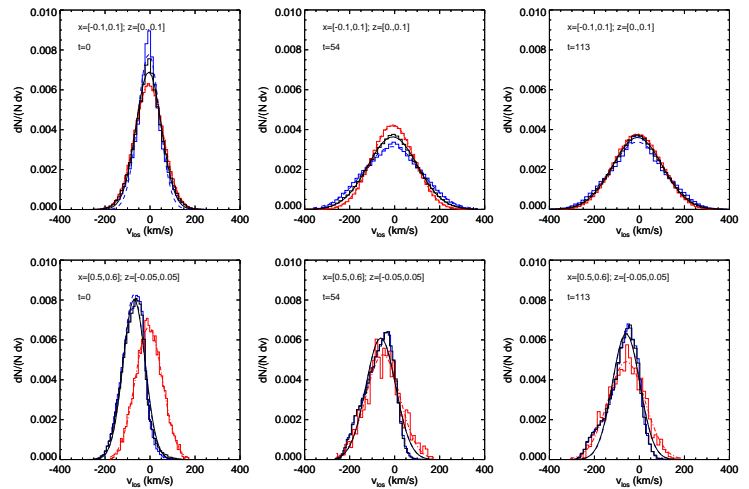}}
\caption{Velocity histogram along the minor (upper panels) and major axis (lower panels) 
of the galaxy model. There is no particular signature of the classical bulge component 
that resides inside the boxy/peanut bulge. Each histogram is fitted with a 
Gaussian distribution. Blue and red histograms are due to disk and classical bulge 
stars alone. The solid black lines represent the histograms of the composite bulge.}
\label{fig:velhist}
\end{figure}

We then examined the velocity histograms at different locations within the 
B/P bulge region, more specifically within $R_{bd}=0.83 R_d$. First, we looked 
at the velocity histograms along the minor axis of the bulge as shown in 
Fig.~\ref{fig:velhist} at three
different epochs during the secular evolution. Each figure (in both the panels) consists
of $3$ histograms : for classical bulge stars alone, disk stars alone and the
composite bulge (classical + B/P bulge) stars. Note that such an exercise
is possible in our N-body simulations as we have a unique ID attached to each
particle. As often the case, the bulge, whether classical, B/P or composite bulge, do
not show any net rotation about the minor axis. As time progresses, the bar
heat up the stars \citep{Sahaetal2010} as a result of which all the histograms fatten. 
Interestingly, at $t=113$, all the histograms
(due to the classical bulge stars, B/P bulge stars and of the composite bulge) are
nearly identical; in other words the minor axis histograms bear no distinguishable
sign of the classical bulge.
Next we looked at a small area ($x/R_d=[0.5, 0.6], |y/R_d|=0.05$) along the major 
axis of the composite bulge - this is depicted in the lower panel of the figure.
Initially, the classical bulge stars show no net rotation (histograms in red)
and the disk stars show net rotation as per construction of the initial model. 
At $t=54$ and $113$, the classical bulge stars show net rotation (see non-zero mean 
of the velocity histograms) and this nearly coincides with the B/P bulge as well as
the composite bulge. Although there is a separate rotating stellar component
(here, the classical bulge) sitting inside the B/P bulge, the net velocity histograms
of the composite bulge is almost identical to that of the B/P bulge and no detectable signatures
of the rotating classical bulge.

At this point, it is useful to recall that the density structures of the classical bulge
shows that there is a bar (see Fig.~\ref{fig:contdenmap} in the central region. 
This was referred to as the classical bulge-bar, reported in \cite{Sahaetal2012}. 
This classical bulge-bar rotates in sync with the disk bar as can be inferred from
position angle of the two of bars at different times during the evolution. 
As a result of this, there is a fine blend of the classical bulge
stars and disk stars that are part of the bar/B/P bulge, explaining the
near identicalness of the velocity histograms. In other words, this would explain why 
the central region, although a composite bulge, behaves both 
morphologically and kinematically like a B/P bulge with a perfectly hidden 
low-mass classical bulge. 

\section{Discussion and Conclusions}
\label{sec:discussion}
It is generally believed that spheroidal stellar system, of which classical bulges
are a subset, do not take take part in the secular evolution.
Indeed, this may be true if such classical bulges are massive and bigger. But it has
 been shown by \cite{Sahaetal2012} that very low mass classical bulges
get substantially affected by the secular evolution driven by a bar. 
In fact, what bar driven secular evolution does to a cool stellar disk is to produce a central
concentration \citep{KK2004}, essentially growing a hotter bulge with a $V/\sigma$ less than that
of the disk. Kinematically, a B/P bulge is essentially a vertically thickened bar with 
hotter stars than what lies beyond the bar's corotation. What secular evolution did to this low
mass classical bulge is substantial and quite similar to the disk. The central part of 
this classical bulge grew a bar \citep{Sahaetal2012} that is nearly identical to the disk bar. 
The bar later heat 
the bulge stars as is evident from the minor axis velocity histograms. From this point of
view, the classical bulge went through a secular evolution parallel to the disk with similar 
end products, although not quite the same. The generality of this outcome is not clear at 
the moment as it requires examining a large number of such high resolution 
N-body simulations and will be the subject matter of a future paper in the series.
In any case, the current analysis produces a twist in a sense that {\it an apparently 
pure B/P bulge could actually be a composite bulge with photometrically and kinematically 
hidden low-mass classical bulge} which could possibly have their origin in mergers/minor mergers 
\citep{Khochfar2006} in accordance with $\Lambda$CDM paradigm of galaxy formation.

The next obvious question is whether the presence of these very low mass classical bulges 
can ever be established unambiguously. Observationally, classical bulges tend to show 
older stellar population,
quiescent in star formation activity and populate, in general, the red-sequence
of the color-magnitude diagram. In contrast, pseudobulges/B/P bulges exhibit intense star
formation and tend to occupy the blue cloud of the color-magnitude diagram. Interestingly,
galaxies with comparatively lower mass classical bulges show some amount of
star formation activity and about half of these galaxies are barred \citep{Gadotti2009}. 
In other words, these barred galaxies would be hosting composite bulges with star 
formation activity intermediate between the pseudobulges and the ones hosting massive 
classical bulges. If these star formation is driven by minor mergers 
\citep{Kaviraj2014,Sachdevaetal2015}, the preexisting classical bulge would further 
be contaminated and might increase the level of complexity for the stellar population
analysis to draw a firm conclusion on the evidence of a pristine classical bulge.
Metallicity gradient could be an important diagnostic to argue for the evidence
of a classical bulge, as it has been done for our Milky Way 
\citep{zoccalietal2008,Gonzalezetal2013,Johnsonetal2013}. Even this can turn out to be a 
difficult task as the bar would act as a strong agent for mixing and 
migration \citep{SellwoodBinney2002} which would affect the 
classical bulge stars, might as well erase its imprint. However, if these bulges are formed
at the very early epoch of the galaxy assembly history, they might contain extreme 
metal poor stars like those found in the Milky Way's bulge \citep{GarciaPerezetal2013,Howesetal2014} and might convey an important message about the presence of an old classical bulge; of course 
a number of possibility remains as they could be the halo stars or thick disk stars. In the end,
one perhaps has to rely on chemodynamical analysis to obtain a reliable answer.
\vspace{-0.1cm}          
\section*{Acknowledgement}
The author thanks the anonymous referee for useful suggestions which improved the manuscript.


\end{document}